\documentclass[aps,prb,superscriptaddress,
amsfonts,amsmath,amssymb,bm,twocolumn
,showpacs
]{revtex4}
\usepackage{graphicx}

\begin{document}
\flushbottom 

\title{Elasticity-driven interaction between vortices in type-II superconductors}

\author{A. Cano}
\author{A. P. Levanyuk}
\affiliation{Departamento de F\'\i sica de la Materia Condensada, C-III, Universidad Aut\'onoma de Madrid, \\ E-28049 Madrid, Spain}
\author{S. A. Minyukov}
\affiliation{Institute of Crystallography, Russian Academy of Sciences,
Leninskii Prospect 59, Moscow 117333, Russia }

\date{ \today}

\begin{abstract}
The contribution to the vortex lattice energy which is due to the vortex-induced strains is calculated covering all the magnetic field range which defines the vortex state. This contribution is compared with previously reported ones what shows that, in the most part of the vortex state, it has been notably underestimated until now. The reason of such underestimation is the assumption that only the vortex cores induce strains. In contrast to what is generally assumed, both core and non-core regions are important sources of strains in high-$\kappa$ superconductors. 

\end{abstract}


\maketitle

\section{Introduction}

Since long time ago, much attention has been paid to the role of long-range strain fields in the vortex state of type-II superconductors. It is well known, for instance, that interaction between defect-induced strains and vortices causes pinning phenomena. These phenomena have been extensively studied almost since Abrikosov predicted the superconducting vortices (see, e.g., Refs.\cite{Anderson62,Kramer67,Galaiko68,Labush68,Miyahara69,Campbell72,Kronmuller73,Chudnovsky01}). It is also known that vortex-induced strains give a contribution to the energies of the vortex lattices (VL). This contribution proves to be essential when discussing the observed correlations\cite{observed_correlation} between VL's and crystal lattices in anisotropic superconductors.\cite{Ullmaier73,Kogan95,Miranovic95} The vortex-induced strains might be important in vortex inertia also, because they contribute to the effective masses of vortices.\cite{Simanek91}

In this paper, we calculate the contribution to the VL energy which is due to the vortex-induced strains. Comparison with the previously reported calculations\cite{Ullmaier73,Kogan95,Miranovic95} shows that, for magnetic fields not so close to the upper critical field $H_{c2}$, this contribution has been notably underestimated until now. The reason of such underestimation is connected with the fact that, contrary to what is assumed in many occasions, the vortex core is not the primary source of strain when the Ginzburg-Landau parameter $\kappa$ of the superconductor is large. 

To clarify this point we shall revise, first of all, the strain induced by a single vortex. This strain is due to all the spatial variations of the density of superconducting electrons that the vortex provokes. The vortex core is a region of strong variations, but is not the only one. There also exist a region of smooth variations which is associated with the presence of superconducting currents. In high-$\kappa$ superconductors, the size of the latter region is much larger than that of the core. Just because of this greater extension, the non-core variations of the density of superconducting electrons are what finally emerge as the main sources of strains.  

Previous calculations\cite{Ullmaier73,Kogan95,Miranovic95} of the elasticity-driven interaction between vortices was based on models that assume, from the beginning, that only the vortex cores induce strains. So an important source of strains in high-$\kappa$ superconductors is overlooked in all these works. But note that, even doing so, it was shown that this interaction was strong enough to explain the observed correlations between VL's and crystal lattices in $\rm NbSe_2$. We revise this elasticity-driven interaction showing that the proper inclusion of all the sources of strain increases its importance in the corresponding problems. 

Let us mention that we evaluate this interaction taking into account all the elastic degrees of freedom of free samples of finite size, i.e. taking into account that both homogeneous and inhomogeneous deformations are possible. In a general case, the elasticity-driven interaction between vortices include contributions due to both homogeneous and inhomogeneous deformations. In the elastically isotropic case the contribution due to the inhomogeneous deformations vanishes. But in any case, the order of magnitude of the total interaction coincides with that of the contribution which is due to the homogeneous deformations. 

The consideration of homogeneous deformations provides us, in addition, an useful technical trick. It is based on to evaluate first the VL energy in the case in which the superconductor is elastically isotropic and its shear modulus $\mu $ is infinite. In this case, the calculations are free of approximations and are almost trivial. If $\mu =\infty $, the only elastic degree of freedom of the sample is its homogeneous dilatation. Therefore, using already known formulas for the VL energy and taking into account the dependence on the dilatation of the corresponding coefficients, the elastic contribution can be easily computed. As we shall see, any isotropic case can be reproduced from this $\mu =\infty $ one. Moreover, the previously reported results can be easily checked by evaluating them for $\mu =\infty$ and comparing them with those obtained considering this case from the beginning. 

Let us mention also that we use the Fourier method when computing the VL energy in the elastically anisotropic case. This method permits to satisfy quite easily the boundary conditions of the elastic problem which correspond with those that take place in real experiments. Thus, one avoids to reproduce spurious effects that a lack of attention to these conditions might give. One of such effect is, for instance, the sample form dependence of the elasticity-driven interaction between vortices (the same group of authors reported this dependence in Ref.\cite{Kogan95} but not in Ref.\cite{Miranovic95}). 

\section{On the elastic effects within the London limit} 

When studying the influence of the elasticity on the vortex properties, many authors use an assumption which might seem quite natural (see, e.g., Refs.\cite{Kramer67,Miyahara69,Ullmaier73,Kogan95,Miranovic95}). It consists of using the ``London approximation'' introduced by Abrikosov in Ref. \cite{Abrikosov57} (see also Ref. \cite{Abrikosov}). However, the essence of this approximation could easily be misinterpreted. As it is frequently commented, within the London approximation the order parameter modulus varies significantly inside of the vortex cores only. Since the spontaneous deformation associated with the superconductivity is proportional to square of the order parameter modulus, it seemed natural that only the core regions ($\rho \lesssim \xi$) are essential sources of stresses. It is just what is assumed in Refs. \cite{Kramer67,Miyahara69,Ullmaier73,Kogan95,Miranovic95}. However, one has to bear in mind that supercurrents also produce an elastic effect because they diminish the value of the order parameter modulus. Locally this diminishing is small. But since the supercurrents occupy a very broad region ($\rho \lesssim \lambda_L$), their effect might be comparable and even more important, as virtually proves to be, than that of the cores.

To make this point more clear, let us recall how the vortex self-energy per unit length $\varepsilon_0$ is calculated within the London limit.\cite{Abrikosov57,Abrikosov} Within this limit one assumes that, when calculating the supervelocity $v_{s}$ from the Ginzburg-Landau equations, the density of superconducting electrons (the square of the order parameter modulus $f^2$) is constant in the corresponding equation. This makes it possible to find out explicitly the spatial distribution of the supervelocity. After doing so, one can follow two different ways:
\begin{itemize}
\item[(a)] Following de Gennes,\cite{de Gennes64} the vortex self-energy is presented as a sum of the magnetic field energy and the kinetic energy of the superconducting electrons:
\begin{align}
\varepsilon_0=\int \left(H^{2}+f^2v_s^2 \right)d^2\boldsymbol{\rho}
\end{align}
[we use here the reduced units, see Ref. \cite{Abrikosov57,Abrikosov}, which are analogous of those defined in Eqs. \eqref{dimless} (see below)]. Integration is carried out taking into account the already found supervelocity, and considering that the density of superconducting electrons is constant. This approximation is justified by virtue of the high value of $\kappa$: $f^2$ diminishes significantly only at $\rho \lesssim \xi $, whereas $v_s^2$ does at $\rho \gtrsim \lambda_L$.

\item[(b)] Following Abrikosov,\cite{Abrikosov57,Abrikosov} the vortex self energy is calculated from the exact formula 
\begin{align}
\varepsilon_0 = \int \left[H^{2}+{1\over 2}\left( 1-f ^{4}\right)\right]d^{2}\boldsymbol{\rho}  
\label{Eq1}\end{align}
(as before, we use here dimensionless quantities). The principal part of this integral arises from the second term, and it is associated with distances much larger than $\xi $. In other words, those variations of $f$ that takes place out of the vortex core are now essential. 
\end{itemize}
As we see, to assume that within the London approximation $f$ is constant out of the vortex cores is not always correct. But, as we have pointed out, this is just the assumption that unfortunately many authors made. For example, when studying the interaction between vortices and lattice defects, Miyahara {\it et al.}\cite{Miyahara69} considered integrals which are similar to Eq. \eqref{Eq1} but, at the same time, neglected all the spatial variations of $f$ at $\rho \gtrsim \xi $. 

It is quite surprising that this assumption has not been critically revised up to now, especially by noting that, in principle, the importance of the out-of-core region for the elastic effects could be understood since long ago. Galaiko\cite{Galaiko68} considered the interaction between vortices and dislocation-induced strains. He found that this interaction depends not only on $\xi $, but also on $\lambda_L $. However, he did not comment Ref. \cite{Kramer67} and discussed neither the vortex-induced strain nor the strain-induced interaction between vortices. Ref. \cite{Chudnovsky01} is a recent example in which the out-of-core region is taken into account when studying an elasticity related problem: the structure of a superconducting vortex pinned by a screw dislocation.

\section{One Single Vortex \label{single_vortex}} 

\subsection{Vortex-induced strain\label{vortex_strain}}

Let us proceed with the calculation of the strain field induced by one single vortex. When doing so, we shall account for all the spatial variations, core and non-core ones, that are associated with the vortex. 

The free energy can be presented as 
\begin{align}
F=F_1+F_2={1\over v}\int (\mathcal{F}_1+\mathcal{F}_2)dv,
\label{F}\end{align} 
where $v$ is the volume of the system, and 
\begin{subequations}
\begin{align}
\mathcal{F}_1\negthickspace&=\negthickspace
 {H^2 \over 8 \pi }  + \negthinspace
 a\left| \Psi \right|^2\negthickspace 
 +{b\over 2}\left| \Psi\right|^4 \negthickspace
+\negthickspace{1\over 4 m}\left|\negthickspace
  \Big( \negthickspace -i\hbar  \nabla \negthickspace - \negthickspace{2e\over c}\mathbf{A} \Big)
                           \Psi\right|^2\negthickspace,
\label{F1}\\
\mathcal{F}_2&=
\alpha_{ij}
\left|\Psi\right|^2 u_{ij}
 +{1\over 2}\lambda_{ijkl}u_{ij}u_{kl}.
\label{F2}
\end{align}\end{subequations} 
Here and below, summation over double indices is implied.

The equations of equilibrium read\cite{Abrikosov,Landau}
\begin{subequations}\label{EE}\begin{gather}
\left[a+b|\Psi|^2\negthickspace + \alpha_{ij}u_{ij}+{1\over 4m }\Big(\negthickspace-i\hbar\nabla \negthickspace- {2e\over c}\mathbf{A}\Big)^2\right]\Psi =0,
\label{GL1}\\
\mathbf{\nabla}\times \mathbf{H} =
{4\pi e\over mc}\left[
{\hbar \over 2i}(\Psi^*\nabla \Psi - \Psi\nabla \Psi^*)
-{2e\over c}|\Psi|^2\mathbf{A}\right], \label{GL2}\\
\lambda_{ijkl}\langle u_{kl}\rangle +\alpha_{ij}\langle |\Psi|^2\rangle=0,\label{Elast_h} \\
{\partial \over \partial x_j}\left(\lambda_{ijkl}u_{kl}+\alpha_{ij}|\Psi|^2\right) =0,\label{Elast_nh} \end{gather}\end{subequations}
where $\langle \dots \rangle $ means volume average. We shall look for the solution of these equations for the case of a single vortex. The $z$-axis of the coordinate frame we choose is parallel to the vortex. The crystal frame is obtained from this coordinate frame by rotation. 

It is clear that far enough from the vortex both the order parameter and the strain tensor tend to constant values; say $\Psi_s$ and $u_{ij}^s$ respectively. Assuming that $\langle |\Psi|^2\rangle\simeq |\Psi_s|^2$, the equations of equilibrium reduce to 
\begin{subequations}\begin{gather}
a + b|\Psi_s|^2 + \alpha_{ij}u_{ij}^s=0,\label{}\\
\lambda_{ijkl}u_{kl}^s+\alpha_{ij}|\Psi_s|^2=0.\label{}\end{gather}\end{subequations}
In consequence: 
\begin{gather}
|\Psi_s|^2=-{a/b^*},\label{psi_s}\\
u_{ij}^s=a\alpha_{kl}\lambda_{ijkl}^{-1}/b^*,
\end{gather}
where $b^*=b-\alpha_{ij}\alpha_{kl}\lambda_{ijkl}^{-1}$ ($\lambda_{ijkl}^{-1}$ is given by $\lambda_{ijkl}^{-1}\lambda_{ijk'l'}=\delta_{kk'}\delta_{ll'}$). These values are just what one obtains in the superconducting phase.

Putting $u_{ij}= u_{ij}^s+ u_{ij}^{v}$,
we can rewrite the equation of equilibrium \eqref{GL1} as 
\begin{align}
\Bigg[1-{|\Psi|^2\over |\Psi_s|^2}- 
{\alpha_{ij}u_{ij}^v\over |\Psi_s|^2b}
+ \xi^2\Big(\nabla - {2ie\over \hbar c}\mathbf{A}\Big)^2\Bigg]\Psi =0, 
\end{align}
where $\xi^2={\hbar^2/(4m|\Psi_s|^2b)}$. It is convenient to introduce the following notation:
\begin{align}
\begin{matrix}
\lambda_L=\sqrt{\displaystyle mc^2 \over \displaystyle 8\pi e |\Psi_s|^2},&&
H_c={\displaystyle \hbar c \over \displaystyle 2\sqrt 2 e \xi \lambda_L}\\ \\
\Psi' ={\displaystyle \Psi\over \displaystyle \Psi_s},&&\mathbf{r'}={\displaystyle \mathbf{r}\over \displaystyle \lambda_L},\\ \\
\mathbf{H'}={\displaystyle \mathbf{H}\over \displaystyle \sqrt 2 H_c},&&\mathbf{A'}={\displaystyle \mathbf{A}\over \displaystyle \sqrt 2 H_c \lambda_L},\\ \\
\hat \alpha'={\displaystyle \hat \alpha\over \displaystyle |\Psi_s|^2b},&&
\hat \lambda'={\displaystyle \hat \lambda\over \displaystyle |\Psi_s|^4b}.
\end{matrix}
\label{dimless}\end{align}
Thus, the equations of equilibrium can be written as (we omit primes in the new quantities)
\begin{subequations}\begin{gather}
(1-v_s^2 -\alpha_{ij}u_{ij}^v)f - f^3=-\kappa^{-2}\triangle f , \label{GL1'_fv}\\
\mathbf{\nabla}\times \mathbf{H} =
\mathbf{v}_sf^2, \label{GL2'_fv}\\
\lambda_{ijkl}\langle u_{kl}\rangle +\alpha_{ij}\langle f^2\rangle=0,\label{Elast_h'} \\
{\partial \over \partial x_j}\left(\lambda_{ijkl}u_{kl}+\alpha_{ij}f^2\right) =0,\label{Elast_nh'} 
\end{gather}\end{subequations}
where the order parameter has been expressed as $\Psi=fe^{i\chi}$, with $\mathbf{v}_s=\kappa^{-1}\nabla\chi - \mathbf{A}$ the above mentioned supervelocity. Here $\kappa = \lambda_L/\xi$ represents the Ginzburg-Landau parameter in our case, which does not differ substantially from the conventional one ($b^* \simeq b$).

The spatial distribution of the supervelocity $\mathbf{v}_s$ can be obtained from Eq. \eqref{GL2'_fv} by assuming that $f$ is constant there, i.e. within the London limit. Thus one finds that $v_s=\kappa^{-1}K_1(\rho)$, where $K_1$ is the MacDonald function (see, e.g., Ref.\cite{Abrikosov}).

In Eq. \eqref{GL1'_fv}, the term with $u_{ij}^v$ results to be of order $\hat \alpha ^2$ because of Eqs. \eqref{Elast_h'} and \eqref{Elast_nh'}. Since $\hat \alpha$ is small,\cite{nota1} the vortex-induced strain can be calculated to the lowest order in $\hat \alpha$ neglecting the changes in $f$ that the term $u_{ij}^v$ in Eq. \eqref{GL1'_fv} induces. In other words, $f^2$ in Eqs. \eqref{Elast_h'} and \eqref{Elast_nh'} can be taken as the solution of Eq. \eqref{GL1'_fv} with $\hat \alpha=0$. This solution can be written as $f^2=1-h$, where $h$ represents the vortex contribution. Using the same approximation that in Ref. \cite{Abrikosov} we have  
\begin{align}
h(\boldsymbol\rho)=\begin{cases}
v_s^2(\rho)& \rho \gg \kappa^{-1},\\
1-C(\kappa \rho)^2,  & \rho \ll \kappa^{-1},
\end{cases}\label{h_r}
\end{align}
where $C$ is a constant of order unity. 

We present the vortex induced strain as\cite{Larkin_Pikin}
\begin{align}
u_{ij}^v&=\epsilon_{ij} + {i\over 2}\sum_{\mathbf{q} \not = 0}\left[q_iu_j(\mathbf{q}) + q_ju_i(\mathbf{q}) \right]e^{i\mathbf{q}\cdot \boldsymbol{\rho}} \nonumber \\
&\simeq\epsilon_{ij} + {iA\over 8\pi^2}
\int d^2\mathbf{q}\left[q_iu_j(\mathbf{q}) + q_ju_i(\mathbf{q}) \right]e^{i\mathbf{q}\cdot \boldsymbol{\rho}}, 
\label{u_LP}\end{align}
where $A$ represents the section of the sample in perpendicular to the vortex (recall that we have split the total strain into $u_{ij}^s +u_{ij}^v$). Here $\epsilon_{ij}$ accounts for the homogeneous deformations that the vortex induces, and $u_i (\mathbf{q})$ is the $i$-th component of the displacement vector in Fourier space. Thus, Eqs. \eqref{Elast_h'} and \eqref{Elast_nh'} can be written as 
\begin{subequations}\begin{gather}
\lambda_{ijkl}\epsilon_{kl}-\alpha_{ij}\langle h\rangle=0,\label{Elast_h} \\G_{ik}^{-1}(\mathbf{q}) u_{k}(\mathbf{q})+iS_i(\mathbf{q})h(\mathbf{q}) =0.\label{Elast_nh} \end{gather}\end{subequations}
where $S_i(\mathbf{q})=\alpha_{ij}q_j$, $G_{ik}^{-1}(\mathbf{q})=\lambda_{ijkl}q_jq_l$, and $h(\mathbf{q})$ is the Fourier transform of the function \eqref{h_r}. For the strain field we have:
\begin{subequations}\begin{align}
\epsilon_{ij}&=\alpha_{kl}\lambda_{ijkl}^{-1}
\langle h\rangle, 
\label{epsilon_ij}\\
u_{i}(\mathbf{q})&=
-i S_k(\mathbf{q})G_{ki}(\mathbf{q})h(\mathbf{q}).
\end{align}\end{subequations}

When calculating the strain tensor at a fixed distance $\rho$ from the vortex [see Eq. \eqref{u_LP}], the inhomogeneous deformations are mainly given by those terms with $q\ll \rho^{-1}$. So the main contribution at $\rho\gg 1$ arises from $q\ll 1$. For these small $\mathbf{q}$'s, the function $h(\mathbf{q})$ can be split into core and non-core contributions:
\begin{subequations}\label{f_q}
\begin{align}
h_{\rm core}(\mathbf{q})\simeq &
{1\over A}\int_{0}^{\kappa^{-1}}\negthickspace\negthickspace\int_{0}^{2\pi}(\rho-\kappa^2\rho^3)
e^{-iq\rho\cos\theta}d\rho d\theta
\nonumber \\=&
{2\pi \over A}\int_{0}^{\kappa^{-1}}
\negthickspace\negthickspace
(\rho -\kappa^2\rho^3)J_0(q\rho)d\rho 
\underset{q \ll 1}{=}{\pi \over 2A\kappa^2}
,\label{f_core}\end{align}
\begin{align}
h_{\rm non-core}(\mathbf{q})\simeq&
{1\over A \kappa^2 }\int_{\kappa^{-1}}^{1}\int_{0}^{2\pi}
\rho^{-1}{e^{-iq\rho\cos\theta}}d\rho d\theta
\nonumber \\=&
{2\pi\over A\kappa^{2} }\int_{\kappa^{-1}}^{1}
{J_0(q\rho)\over \rho}d\rho
\underset{q \ll 1}{=}
{2\pi \over A \kappa^2}
\ln \kappa 
\label{f_non-core}\end{align}\end{subequations}
(here we have used the asymptotic form of $v_s\approx 1/(\kappa \rho)$ for $\kappa^{-1}\ll \rho \ll 1$, see Ref. \cite{Abrikosov}).

As a result, at $\rho \gg 1$ the strain tensor can be written as
\begin{align}
u_{ij}^v(\boldsymbol{\rho})&=\eta
\Bigg[{\alpha_{kl}\lambda_{ijkl}^{-1}\over A}
+\negthickspace\int\negthickspace
{d^2\mathbf{q}\over (2\pi)^2}
q_iS_k(\mathbf{q})G_{kj}(\mathbf{q})
e^{i\mathbf{q}\cdot \boldsymbol{\rho}}\Bigg]\nonumber \\
&=\eta
\Bigg[{\alpha_{kl}\lambda_{ijkl}^{-1}\over A}
+{1\over \rho^2}\int_0^{2\pi}\Theta_{ij}(\theta_{\mathbf{q}})d\theta_{\mathbf{q}}
\Bigg]\label{u_ij(r)},
\end{align}
where $\eta=\int h(\rho) d^2\boldsymbol{\rho}=\pi(1 + 4\ln\kappa)/(2\kappa^{2})$, and $\Theta_{ij}$ is a tensor which depends only on the angle $ \theta_{\mathbf{q}}$ ( $\mathbf{q}\cdot \boldsymbol{\rho}=q\rho\cos\theta_\mathbf{q}$). If the sample is large enough the first term in Eq. \eqref{u_ij(r)} can be neglected. But we retain it because, when dealing with the strain-induced interaction (see below), its contribution becomes significant (this fact is well known in the theory of point defects, see e.g. Refs. \cite{Khachaturyan}). Note that the non-core contribution to $\eta$, i.e. the logarithmic term, could also be obtained from the well known expression of the vortex self-energy: according to Abrikosov,\cite{Abrikosov}  $\varepsilon_0 \simeq {1\over 2} \int (1 - f^4)d^2\boldsymbol{\rho }\simeq 2\pi \int_{\kappa^{-1}}^1h(\rho)\rho d\rho=2\pi
\kappa^{-2}\ln \kappa $.

Kogan {\it et al.}\cite{Kogan95} obtained a similar expression for vortex-induced strain considering an infinite medium. In such a case, the first term of Eq.\eqref{u_ij(r)} vanishes at all. But the main difference between Eq. \eqref{u_ij(r)} and the expression reported by Kogan {\it et al.}\cite{Kogan95} resides in the corresponding values of $\eta$. Assuming that only the vortex core induces strain, Kogan {\it et al.} reported a value $\pi /\kappa^{2}$. So they overlooked the logarithmic term in $\eta=\pi(1 + 4\ln\kappa)/(2\kappa^{2})$ that arises because of non-core contributions. This implies that in the case of high-$\kappa$ superconductors, Kogan {\it et al.} strongly underestimated the vortex-induced strain. 

\subsection{Elasticity-driven interaction between vortices: Qualitative estimations\label{Q_estim}}

Let us now estimate the interaction energy of a VL which is associated with the strains that the vortices induce. As we have pointed out before, the inhomogeneous part of these strains have been previously reported but neglecting non-core contributions (see, e.g., Ref. \cite{Kogan95}). If the distance between vortices is much longer than $\lambda_L$, to take into account these non-core contributions reduces to modify the strains by a factor. In consequence, the interaction energy one obtains by taking into account both core and non-core contributions coincides, up to the corresponding factor, with previously reported ones. Kogan {\it et al.},\cite{Kogan95} for instance, evaluated the interaction energy of a VL by summing up all pairwise contributions. Modifying this interaction energy by including the non-core contributions, one can see that 
\begin{align}
F_{\rm int}^{\rm (nh)}\sim -{(1+4\ln\kappa)^2\over \kappa^2}{\Delta K\over K}B^2.
\label{F_int^nh}\end{align}
Here $\Delta K/K$ stands for the order of magnitude of the relative change in the elastic moduli due to the normal-superconducting transition, and $B$ represents the magnetic induction.

The interaction between vortices that arise due to the homogeneous strains can be easily estimated as follows. 
It is clear that $N$ vortices will induce a total (homogeneous) strain $N\epsilon$, where $\epsilon$ is given by Eq. \eqref{epsilon_ij}, if the distance between them is large enough. When substituting this strain in the corresponding terms of VL energy: $-\alpha (N\epsilon)(N\langle h \rangle) + \lambda (N \epsilon)^2/2$, one obtains $-n^2 \eta^2 \alpha^2 /(2\lambda) $, where $n=N/A$ is the vortex density. This is precisely the interaction term that we are looking for. Taking into account that the vortex density is $n=\kappa B/(2\pi)$, and $\alpha^2/\lambda= \Delta K/K$ [recall that we are using the dimensionless units defined in Eq. \eqref{dimless}]; this interaction can be estimated as
\begin{align}
F_{\rm int}^{\rm (h)}\sim -{(1+4\ln\kappa)^2\over \kappa^2}{\Delta K\over K}B^2.
\label{F_int^h}\end{align}

As we see, the order of magnitude of both interaction terms Eqs. \eqref{F_int^nh} and \eqref{F_int^h} coincide. Consequently, either of them give us an estimate of the order of magnitude of the total interaction energy.

\section{Vortex lattice: Elastically isotropic medium \label{isotrooopic}}

It is convenient to begin the treatment of VL's considering the case elastically isotropic superconductors. In this case, the elastic contribution to the VL energy can be obtained, without any new approximation, from already known formulas for this VL energy. Such formulas are available for the regions $H \approx H_{c1}$, $H\approx H_{c2}$ (Refs. \cite{Abrikosov57,Abrikosov}) and for intermediate fields $H_{c1}\ll H \ll H_{c2}$ (Refs.\cite{Friedel63,de Gennes64}). They reasonably match at the boundaries of the corresponding regions. This permits us to study the elastic effects in isotropic superconductors with the same accuracy. We begin with the case $\mu =\infty $ where the calculations are elemental.

\subsection{Infinite shear modulus \label{mu_infty}}

The only elastic degree of freedom of a system which shear modulus is infinite is its homogeneous dilatation. If the system is not clamped this homogeneous dilatation, say $u$, must be understood as a variational parameter. In the free energy \eqref{F}, this variational parameter modifies the coefficient of the term $|\Psi |^{2}$, which can be rewritten as $a(u)=a+\alpha u$. 

Let us fix the parameter $u$ for a while, i.e., let us consider momentaneously a clamped sample. Thus, after minimizing with respect to all degrees of freedom excepting $u$, the free energy of the VL with respect to that of the superconducting state can be written as a sum of two terms: a $u$-dependent VL energy via the coefficient $a(u)$, and the elastic energy. It is (see, e.g., Ref.\cite{Fetter69} and the references therein)  
\begin{align}
F=F_{\rm VL}(u) +{K \over 2}u^{2},  
\label{freeensum}
\end{align}
where the $F_{\rm VL}$ has the form 
\begin{align}
F_{\rm VL}=
\begin{cases}
{\frac{\displaystyle BH_{c1}}{\displaystyle4\pi }} 
& \text{(I)}, \\ \\
{\frac{\displaystyle1}{\displaystyle8\pi }}\left[ B^{2}+BH_{c1}{\frac{
\displaystyle\ln (\nu d/\xi )^{2}}{\displaystyle\ln \kappa }}\right] 
& \text{(I--II)},\\  \\ 
{\frac{\displaystyle1}{\displaystyle8\pi }}\left[ B^{2}-{\frac{\displaystyle
(H_{c2}-B)^{2}}{\displaystyle1+(2\kappa ^{2}-1)\beta _{A}}}\right] 
& \text{(II)},\\ 
\end{cases}
\label{Londonen}\end{align}
in the corresponding regions of magnetic fields defined as (I): $H\approx H_{c1}$, (I--II): $H_{c1}\ll H\ll H_{c2}$, and (II): $H\approx H_{c2}$.
Here $\beta _{A}=\langle \Psi ^{4}\rangle /\langle \Psi ^{2}\rangle ^{2}=1.16
$ for a triangular VL, and $2\ln \nu =2(\gamma -1)+\ln [\sqrt{3}/(8\pi )]$, where $\gamma (=0.57772\dots)$ is the Euler's constant. The magnetic induction $B$ and the distance between vortices $d$ are such that $B = 2\phi_0/(\sqrt 3 d^2)$ in a triangular VL, where $\phi _0$ is the flux quantum. 

The magnetic induction as a function of the magnetic field is given by:\cite{Fetter69}
\begin{align}
B=\negthickspace
\begin{cases}
{2\phi_0 \over \sqrt 3 \lambda_L^2} \left\{ 
\ln \left[
{3\phi_0\over 4\pi \lambda_L^2(H-H_{c1})}
\right]\right\}^{-2}
& \text{(I)},\\ \\ 
H-H_{c1}+{\phi_0 \over 8\pi \lambda_L^2}
\left\{ 
\ln \left[{4\pi \lambda_L^2(H-H_{c1})\over \phi_0 }
\right]+ \widetilde \gamma \right\}
& \text{(I--II)}, \\ \\ 
H-{H_ {c2} - H \over (2\kappa^{2}-1)\beta_A}
& \text{(II)},
\end{cases}
\label{B(H)}\end{align}
where $\widetilde \gamma = 2(1-\gamma)$. Let us recall that in high-$\kappa$ superconductors one has the following relationships (see, e.g. Refs. \cite{Abrikosov,Fetter69}):
\begin{align}
H_{c1}=&{\ln \kappa\over 2\kappa^2}H_{c2}, 
\label{H_1_2}\\
\left({d \over \xi}\right)^2 =& {4\pi\over \sqrt 3} {H_{c2}\over B}=
{8\pi \over \sqrt 3}{\kappa^2\over \ln \kappa}{H_{c1}\over B}.\label{d_xi}
\end{align}

One can check that at the boundaries of the magnetic field regions, the expressions in Eq. \eqref{Londonen} match one each other with a reasonably accuracy:
\begin{itemize}
\item[$\circ$] For $H\approx \zeta_1 H_{c1}$ ($\zeta_1 \gtrsim 1 $) one has $B \approx \zeta_1 H_{c1}$ according to the expressions (I--II) and (II) in Eq. \eqref{B(H)}. Therefore, taking into account the relationship \eqref{d_xi}, one can obtain the free energy as
\begin{align}
F_{\rm VL}\simeq \begin{cases}
{\displaystyle \zeta_1 H_{c1}^2
\over \displaystyle 4\pi}
& \text{(I)},\\ \\ 
{\displaystyle\zeta_1 H_{c1}^2\over \displaystyle 8\pi}\left(\zeta_1 +{\displaystyle\ln \left[8\pi(\nu\kappa)^2 /(\sqrt 3 \zeta_1 \ln \kappa)\right]\over \displaystyle\ln \kappa}\right)
& \text{(I--II)}.
\end{cases}\end{align}

\item[$\circ$] For $H=\zeta_2 H_{c2}$ ($\zeta_2 \lesssim 1$) one has $B\simeq \zeta_2 H_{c2}$ according to the expressions (I--II) and (II) in Eq. \eqref{B(H)} and the relationship \eqref{H_1_2}. Therefore, taking into account the relationship \eqref{d_xi}, one can obtain the free energy as
\begin{align}
F_{\rm VL}&\simeq \begin{cases}
{\displaystyle  \zeta_2  H_{c2}^2\over \displaystyle  8\pi }\left(\zeta_2  + {\displaystyle
\ln [4\pi \nu ^2/(\sqrt 3\zeta_2)]\over \displaystyle 2\kappa^2}\right)
&\text{(I--II)},\\ \\
{\displaystyle H_{c2}^2\over \displaystyle 8\pi }
\left(\zeta_2^2 - {\displaystyle(1-\zeta_2)^2\over \displaystyle 2\kappa^2} \right)
&\text{(II)}.
\end{cases}
\label{}\end{align}

\end{itemize}

The critical magnetic fields entering all above expressions are $u$-dependent magnitudes:
\begin{gather}
H_{c1}(u)={\frac{\ln \kappa }{\sqrt{2}\kappa }}H_{c}(u)=H_{c1}^{\circ
}+H_{c1}^{\prime }u, \\
H_{c2}(u)=\sqrt{2}\kappa H_{c}(u)=H_{c2}^{\circ }+H_{c2}^{\prime }u,
\end{gather}
where $H_{c}(u)=2a(u)\sqrt{\pi /b}=H_{c}^{\circ }+H_{c}^{\prime }u$ (with $H_{c}^{\circ}=2a\sqrt{\pi /b}$ and $H_{c}^{\prime }=2\alpha \sqrt{\pi /b}$). In consequence, the ratio $d/\xi$ is also a $u$-dependent magnitude [see Eq.\eqref{d_xi}], from which one can write the coherence length $\xi$ as $\xi = \xi^\circ + \xi' u$. Mention that the Ginzburg-Landau parameter $\kappa $ is independent of $u$ because it does not depend explicitly on the coefficient $a(u)$. 

Let us now proceed to minimize the free energy \eqref{freeensum} with respect to $u$, i.e., to take into account that the sample is in fact unclamped. After doing so, we obtain
\begin{align}
F= \begin{cases}
{\frac{\displaystyle1}{\displaystyle8\pi }}
 \left( 2BH_{c1}^{\circ }
 -\delta_{\rm I}B^{2}\right),
 & \text{(I)},\\ \\ 
{\frac{\displaystyle1}{\displaystyle8\pi }}
\left[ B^{2}+BH_{c1}^{\circ }
    {\frac{\displaystyle\ln (\nu d/\xi ^{\circ })^{2}}
    {\displaystyle\ln \kappa}}
    -\delta _{\rm I-II}B^{2}
\right]
&\text{(I--II)},\\ \\
{\frac{\displaystyle1}{\displaystyle8\pi }}
\left[ B^{2}
-{\frac{\displaystyle(H_{c2}^\circ-B)^{2}}
       {\displaystyle 1+(2\kappa^2 -1) \beta _{A}-\beta _{e}}}\right] 
& \text{(II)},
\end{cases}
\label{freeenattr}\end{align}
where 
\begin{align}
\delta _{\rm I}&= 
{\ln ^{2}\kappa \over 2\kappa^2}{\Delta K\over K},
\label{delta_I}\\
\delta _{\rm I-II}&=
{\frac{\left[ 1+2\ln (\nu d/\xi ^{\circ })\right]
^{2}}{16\pi K+{\frac{2\sqrt{2}H_{c}^{\prime 2}B}{\kappa H_{c}^{\circ }}}}}{\frac{H_{c}^{\prime 2}}{\kappa ^{2}}}\simeq 
{\frac{\ln^2 (d / \xi ^{\circ })}
{4\kappa^2}}{\Delta K \over K}
\label{delta_I--II}\\
\beta _{e}&= 
2\kappa ^{2}{\Delta K\over K}.
\end{align}
Here it has been taken into account that 
$H_c^{\prime 2}/(4\pi K)=\Delta K / K$ is the relative change in the bulk modulus due to the normal-superconducting transition. Because this relative change is usually $\Delta K / K \ll 1$, the expression for the region (II) in Eq. \eqref{freeenattr} can be written as
\begin{align}
F\simeq {1 \over 8\pi }
\left[ B^{2}
-{(H_{c2}^\circ-B)^2 \over 1+(2\kappa^2 -1) \beta _{A}}
-\delta_{\rm II}(H_{c2}^\circ-B)^2 \right],
\label{F_en_2}\end{align}
where 
\begin{align}
\delta_{\rm II}= {\beta_e \over 
[1+(2\kappa^2 -1) \beta _{A}]^2
} \simeq {1\over 2\kappa ^2}{\Delta K\over K}.
\label{delta_II}\end{align}

In all above expressions for the free energy, one can identify a term 
\begin{align}
F_{\mathrm{int}}=-{\frac{\delta B^{2}}{8\pi }},
\label{interaction_iso}\end{align}
which describes an attractive interaction between vortices. The coefficient $\delta$ is given by Eq. \eqref{delta_I}, \eqref{delta_I--II} or \eqref{delta_II}, depending on the magnetic field region one considers. It can be presented as  
\begin{align}
\delta \approx {[\zeta +\ln (d/\xi ^{\circ })]^2\over 2\kappa^2}{\Delta K \over K}
\end{align}
taking into account that the ratio $d/\xi ^{\circ }$ must be replaced by $\kappa$ if $d\gtrsim \lambda_L$, and $\zeta $ is a constant of order of unity (see Fig. \ref{F1}). 

According to what we have seen in the preceding section, the logarithmic contribution to the coefficient $\delta$ is due to non-core effects. These effects have been overlooked until now. As we show in Fig.\ref{F1}, the neglection of these non-core effects leads to underestimate the elasticity-driven interaction between vortices. And by virtue of the high value of $\kappa$, such a underestimation is quite significant in almost all the mixed state.  

\begin{figure}[t]
\includegraphics[width=.5\textwidth]{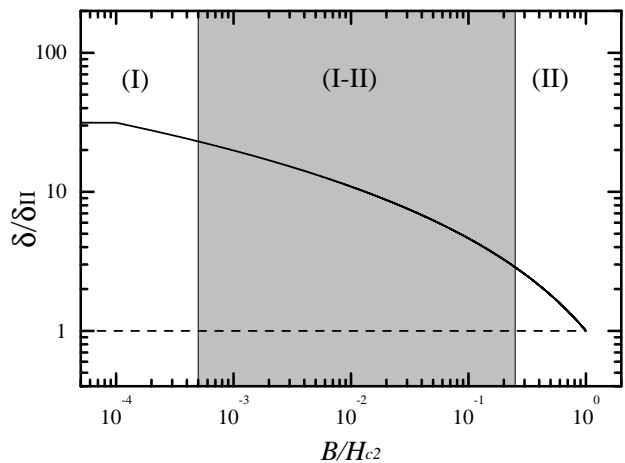}
\caption{Log-log plot of the coefficient $\delta$ of the attraction term $\sim -\delta B^2$ of the free energy as a function of the magnetic induction, taking into account (solid line) and neglecting (dashed line) non-core contributions. The regions indicated as (I), (I-II) and (II) (see text) are the corresponding ones for $\kappa \simeq 100$ (note that $H_{c1}\simeq 10^{-4}H_{c2}$ in this case).}\label{F1}
\end{figure}

\subsection{Finite shear moduli}

It is quite straightforward to extend the results that we have obtained for the $\mu = \infty$ case, to the most general isotropic one. Note that minimizing the free energy \eqref{F} with respect to all elastic degrees of freedom one obtains\begin{align}
F_{2}=-{\frac{\alpha^2}{2K_{4/3}}}
\langle \left| \Psi \right|^{4}\rangle 
-{\frac{\alpha^2}{2K}}{\frac{4\mu }{3K_{4/3}}}\langle \left|
\Psi \right| ^{2}\rangle ^{2},
\label{F_2_nl}\end{align}
where $K_{4/3}=K+4\mu /3$. The first term of this expression renormalizes the coefficient $b$ of Eq. \eqref{F1}. This renormalization disappears in the limit $\mu \to \infty$. The second term makes that the free energy becomes into a non-local functional. This non-locality remains as long as the shear modulus does not vanish.

Working out from this functional one could recover the free energy \eqref{freeenattr} by ascribing the coefficients in Eq. \eqref{F_2_nl} the values they assume in the case $\mu = \infty$, i.e. $\alpha^2/(2K_{4/3})=0$ and $4\mu /(3K_{4/3})=1$. This consideration does not change the functional form of Eq. \eqref{F_2_nl} (only coefficients change). So one concludes that the free energy density of any isotropic type-II superconductor has the form of Eq. \eqref{freeenattr} with the corresponding renormalized constants: 
\begin{align}
b \quad \longrightarrow & \quad b-\alpha^2/K_{4/3}\\
(\alpha^2/K)\quad \longrightarrow &\quad (\alpha^2/K)[4\mu /(3K_{4/3})].
\end{align}
Note that, because the resulting coefficient $\delta$ in Eq. \eqref{interaction_iso} vanishes if $\mu =0$, it can be said that the elasticity-driven interaction between vortices is associated with the solid-state elasticity. 

\subsection{Comparison with previously reported results\label{C_with_R}}

Let us start this section by comparing our results with those reported in Refs. \cite{Ullmaier73,Kogan95}. In these references, intermediate fields far from $T_c$ are considered. Mention that although strictly speaking the Ginzburg-Landau approach that we use is not valid far from $T_c$, it still gives correctly the orders of magnitude. So the comparison still makes sense. Mention also that in Refs.\cite{Ullmaier73,Kogan95} the homogeneous part of the strains are omitted. In Ref. \cite{Ullmaier73} this omission is mentioned explicitly, while in Ref. \cite{Kogan95} it follows from the fact that they consider infinite samples when calculating the interaction between vortex pairs. Therefore, as we argued in Sec. \ref{Q_estim}, we can only compare the order of magnitude (see below for a more detailed comparison). Such a comparison reveals that, as a result of the neglection of the non-core contributions to the interaction energy, this energy is notably underestimated in Refs.\cite{Ullmaier73,Kogan95} through the most part of the mixed state (see Fig. \ref{F1}). Such a underestimation is at least by factor $\sim \ln ^2 \kappa $ close to $H_{c1}$. 

In Ref. \cite{Miranovic95}, treating the case $H\approx H_{c2}$, both homogenous and inhomogeneous strains are seemingly taken into account. In accordance with Eq. (28) of this reference, the free energy in the isotropic case should be of the form 
\begin{align}
F={1 \over 8\pi }
\left[ B^{2}
-{1+(2\kappa^2 -1) \beta _{A} - 4\kappa^2\beta_2\over [1+(2\kappa^2 -1) \beta _{A} + 4\kappa^2\beta_2]^2}(H_{c2}^\circ-B)^2 \right],
\end{align}
where 
\begin{align}
\beta_2=-{\alpha^2 \over (K + {4\over 3}\mu)b }\beta_A.
\end{align}
Because of the smallness of $\beta_2$, this expression can be approximated to 
\begin{align}
F\simeq {1 \over 8\pi }
\left[ B^{2}
-{(H_{c2}^\circ-B)^2 \over 1+(2\kappa^2 -1) \beta _{A}}
-\delta_2 (H_{c2}^\circ-B)^2
\right],
\label{F_kogan}\end{align}
where $\delta_2= -12\kappa^2\beta_2/{[1+(2\kappa^2-1)\beta_A]^2}$.

At first glance, one could think that the last term of this expression and the last one of Eq. \eqref{F_en_2} differ only in a numerical factor. So the elasticity-driven interaction between vortices are reasonably well reproduced by either of them. However, a deeper inspection of Eq. \eqref{F_kogan} reveals that it is erroneous: the coefficient $\delta_2$ (i) vanishes if $\mu=\infty$ and (ii) remains finite if $\mu=0$. Being the strain-driven interaction between vortices due to the specific features of the solid-state elasticity, the results one obtains from Ref. \cite{Miranovic95} in the above mentioned limiting cases cannot be correct. This motivates us to reconsider the problem treated in Ref. \cite{Miranovic95} (Sec. \ref{Hc2} below). 

\section{Vortex lattice: Elastically anisotropic medium} 

Let us reconsider the free energy \eqref{F} and, as usual (see Ref. \cite{Abrikosov}), integrate by parts term with $\nabla \Psi ^*$. Thus, after using the Gauss' theorem and the boundary condition $\mathbf{n}\cdot \left.\big(-i\hbar\nabla-{2e\over c}\mathbf{A} \big)
\Psi\right|_\Sigma=0$ ($\mathbf{n}$ is the unit vector of the normal to the surface $\Sigma$), one finds that:
\begin{align}
{1\over 4 m}\int&\Big|
\Big(-i\hbar\nabla-{2e\over c}\mathbf{A} \Big)
\Psi\Big|^2dv
\nonumber\\ 
&={1\over 4 m}\int \Psi^*
\Big(-i\hbar\nabla-{2e\over c}\mathbf{A} \Big)^2
\Psi dv.
\end{align}
Because $\Psi $ satisfies the equation \eqref{GL1}, this expression can be written as
\begin{align}
{1\over 4 m}\int &\Psi^*
\Big(-i\hbar\nabla-{2e\over c}\mathbf{A} \Big)^2
\Psi dv
\nonumber \\&
=-\int \left(a|\Psi|^2 + b |\Psi|^4 + \alpha_{ij}u_{ij}|\Psi|^2\right)dv.
\end{align}
As a result, the free energy \eqref{F} can be presented as 
\begin{align}
F={1\over v} \int \left({H^2\over 8\pi} - {b\over 2}|\Psi|^4 + {1\over 2}\lambda_{ijkl}u_{ij}u_{kl}\right) dv.
\label{master_F}\end{align}
This expression generalizes the Abrikosov' one [see Eq. \eqref{Eq1}] by taking into account the elastic degrees of freedom. In Ref. \cite{Miranovic95} it has been reported a similar expression that, however, is erroneous (see Sec. \ref{Hc2} below). 

\subsection{$\boldsymbol{H \ll H_{c2}}$}

When treating the fields far from $H_{c2}$, it is convenient to express $|\Psi|^2=|\Psi_s|^2 - h$, where $|\Psi_s|^2 = -a/b^*$ [see Eq. \eqref{psi_s}] and $h$ now represents the VL contribution. In addition, we express the strain tensor as $u_{ij}=u_{ij}^s + u_{ij}^v$, where 
$u_{ij}^s=-\alpha_{kl}\lambda^{-1}_{ijkl}|\Psi_s|^2$ and 
\begin{align}
u_{ij}^v=&\alpha_{kl}\lambda^{-1}_{ijkl}\langle h\rangle
\nonumber \\
&+{1\over 2}\sum_{\mathbf q\not = 0}
\big[q_iS_k(\mathbf{q})G_{kj}(\mathbf{q})
+q_jS_k(\mathbf{q})G_{ki}(\mathbf{q})\big]
h(\mathbf q)e^{i\mathbf{q}\cdot\mathbf{r}}.
\label{e_ij}\end{align}
Thus, the equations of equilibrium \eqref{Elast_h} and \eqref{Elast_nh} are satisfied. Substituting these expressions for $|\Psi|^2$ and $u_{ij}$ into Eq. \eqref{master_F}, one finds that
\begin{align}
F=& F_s
+{1\over v} \int \left(
{H^2\over 8\pi} 
+b^*|\Psi_s|^2 h - {b\over 2}h^2
\right) dv\nonumber \\
&+{1\over 2}\sum_{\mathbf{q}}b'(\mathbf{q})|h(\mathbf q)|^2,
\label{F_rara}\end{align}
where $F_s=-b^*|\Psi_s|^4/2$ and 
\begin{align}
b'(\mathbf{q})=\begin{cases}
\alpha_{ij}\alpha_{kl}\lambda^{-1}_{ijkl}&(\mathbf{q}=0),\\
S_i(\mathbf{q})S_j(\mathbf{q})G_{ji}(\mathbf{q})&(\mathbf{q}\not=0).
\end{cases}
\label{b'}\end{align}
Note that, for $\mathbf{q}\not=0$, the function  $b'(\mathbf{q})$ only depends on the angle.

When calculating $h$, we can retain the lowest order terms in $\hat \alpha $, i.e. we can take $h \simeq h_0 + h_1$ where $h_0$ is given by solving Eq. \eqref{GL1} with $\hat \alpha=0$, and $h_1$ represents the correction to this solution due to the term $\alpha_{ij}u_{ij}^v$ in Eq. \eqref{GL1}:
\begin{align}
bh_1\simeq
\begin{cases}
\alpha_{ij}u_{ij}^v& \text{(out of cores)},\\
0                  & \text{(inside cores)}.
\end{cases}
\label{h_1}\end{align}
The absence of correction inside the cores follows from the fact that, in these regions, Eq. \eqref{GL1} can be linearized because of the smallness of the order parameter modulus (see, e.g., Ref. \cite{Abrikosov}). Let us remark, even doing so, the vortex cores are taken into account: they act as strain sources.

Substituting these expressions in Eq. \eqref{F_rara}, and retaining the lowest order terms, we obtain
\begin{align}
F=& F_s
+b^*|\Psi_s|^2 \zeta \langle h_0 \rangle
+{\langle H^2 \rangle \over 8\pi} 
- {b\over 2}\langle h_0^2\rangle 
\nonumber \\&
-b\sum_{\mathbf q}h_0(-\mathbf{q})h_1(\mathbf{q})
+{1\over 2}\sum_{\mathbf{q}}b'(\mathbf{q})|h_0(\mathbf q)|^2,
\label{F_sum}\end{align}
where $\zeta = 1+b_0'/b$. 

The most important contribution to the two last terms in Eq. \eqref{F_sum} arises from $q<\xi^{-1}$.\cite{nota2} At these $\mathbf{q}$'s, the function $h_1(\mathbf{q})$ can be calculated by taking $bh_1\simeq \alpha_{ij}u_{ij}^v$ in all the regions [see Eq. \eqref{h_1}], so 
\begin{align}
-b\sum_{\mathbf{q}}h_0(-\mathbf{q})h_1(\mathbf{q})\simeq
-\sum_{\mathbf{q}}b'(\mathbf{q})|h_0(\mathbf{q})|^2.
\end{align}
As a result, the free energy \eqref{F_sum} is
\begin{align}
F\simeq & F_s
+b^*|\Psi_s|^2 \zeta \langle h_0 \rangle
+{\langle H^2 \rangle \over 8\pi} 
- {b\over 2}\langle h_0^2\rangle 
\nonumber \\&
-{1\over 2}\sum_{\mathbf{q}}b'(\mathbf{q})|h_0(\mathbf q)|^2.
\label{F_sumq}\end{align}
The last term of this expression represents the strain-induced contribution to the VL energy. The term with $\mathbf{q}=0$ is associated with the homogeneous strains. The elastic constants enter this term through an invariant combination [see Eq. \eqref{b'}], so it does not depend on the orientation of the VL with respect to the crystal axes. This dependence arises from the terms with $\mathbf{q}\not =0$. 

Let us mention that this formula demonstrates that the elasticity-driven interaction between vortices does not depend on the sample form, unlike to the statement made in Ref. \cite{Kogan95}. Indeed, such a dependence would mean
that contribution to the sum from the region of small $q$'s is essential and comparable with the contribution of the rest of the sum. But the function $h_0\left( \mathbf{\rho }\right) -\langle h _0\rangle$ is a periodic function defined in a finite volume (neglecting the near-of-the surface distortions). Its Fourier spectrum does not contain small $q$'s but has maxima at the non-zero reciprocal lattice vectors. The form and the size of the sample is reflected in the form and the width of these
maxima and nowhere else. In fact, the sums over $\mathbf{q}$'s can be replaced by sums over the reciprocal lattice vectors of the VL $\mathbf{Q}$. Putting
\begin{align}
h_0(\boldsymbol{\rho})=\sum_{i}
\widetilde h_0(\boldsymbol{\rho} -\boldsymbol{\rho}_i)
\label{sep}\end{align}
where $\boldsymbol{\rho}_i$ represent the vortex positions, one finds that
\begin{align}
h_0(\mathbf{q})&={1\over A} 
\sum_{i}\int \widetilde h _0
(\boldsymbol{\rho} -\boldsymbol{\rho}_i)
e^{-i\mathbf{q}\cdot\boldsymbol{\rho}}
d^2\boldsymbol{\rho}
= n \widetilde h_0(\mathbf{q}), 
\label{}\end{align}
where $n$ is the vortex density, $A$ represents the section of the sample in perpendicular to the VL, and 
\begin{align}
\widetilde h_0(\mathbf{q}) \equiv
\begin{cases}
\int \widetilde h _0 (\boldsymbol{\rho})
e^{-i\mathbf{q}\cdot \boldsymbol{\rho}_i}
d^2\boldsymbol{\rho}& \text{($\mathbf{q}=\mathbf{Q}$)},\\
0&\text{(otherwise)}, 
\end{cases}
\end{align}
with $\mathbf{Q}$ any of the reciprocal lattice vectors (note that $A^{-1}\sum_i e^{-i\mathbf{q}\cdot \boldsymbol{\rho}_i}=n \delta_{ \mathbf{q}\mathbf{Q}}$). As a result, the last term in Eq. \eqref{F_sumq} can be written as
\begin{align}
F_{\rm el}=-{n^2\over 2}\Big[ b'(0)\widetilde{h}_0^2(0)
+\sum_{\mathbf{Q}\not =0}b'(\mathbf{Q})
|\widetilde{h}_0(\mathbf{Q})|^2\Big].
\label{sum_QQ}\end{align}

Let us emphasize that with this expression, one takes into account that both core and non-core regions act as strain sources. It can be straightforwardly illustrated close to $H_{c1}$. Here, due to the large separation between vortices, the function $\widetilde h_0$ in Eq. \eqref{sep} practically coincides with that associated with one single vortex [see Eq. \eqref{h_r}]. In consequence the function $\widetilde h_0(\mathbf{Q})$ varies slowly up to $Q \approx \xi^{-1}$ and then rapidly drops to zero. So in Eq. \eqref{sum_QQ} it can be taken as $\widetilde{h}_0(\mathbf{Q})\simeq 
\widetilde{h}_0(0)$, which naturally split into core and non-core contributions [see Eqs. \eqref{f_q}]:
\begin{align}
\widetilde{h}_0(0)=\pi(1 + 4\ln\kappa)|\Psi_s|^2\xi^2/2,
\label{}\end{align}
limiting the sum over ${\mathbf{Q}}$'s up to $Q_{\rm max} \lesssim \xi^{-1}$. 

Let us calculate $F _{\rm el} $ explicitly in the isotropic case. In this case one has $\alpha_{ij}=\alpha\delta_{ij}$ and $\lambda_{ijkl}=(K-{2\over 3}\mu)\delta_{ij}\delta_{kl} + \mu (\delta_{ik}\delta_{jl}+\delta_{il}\delta_{jk})$, where $K$ and $\mu$ are the bulk and the shear modulus respectively. In consequence,
\begin{align}
b'(\mathbf{Q})=\begin{cases}
\alpha^2/K&(\mathbf{Q}=0),\\
\alpha ^2 / (K + 4\mu/3)&(\mathbf{Q}\not=0),
\end{cases}
\label{b'_iso}\end{align}
and Eq. \eqref{sum_QQ} yields
\begin{align}
F_{\rm el}&=-{n^2\over 2}\widetilde{h}_0^2(0)
\Bigg(
{\alpha^2\over K} 
+\sum_{\mathbf{Q}\not =0}^{\mathbf{Q}_{\rm max}}
{\alpha ^2 \over  K + {4\over 3}\mu}
\Bigg)\nonumber \\
&=-{n^2\over 2}\widetilde{h}_0^2(0)
\Bigg[
{4\alpha^2\mu \over 3K(K+{4\over 3}\mu)} 
+\sum_{\mathbf{Q}=0}^{\mathbf{Q}_{\rm max}}
{\alpha ^2 \over  K + {4\over 3}\mu}
\Bigg]
\nonumber \\& \simeq 
-{\alpha^2\widetilde{h}_0^2(0)\over 2(K+{4\over 3}\mu)}\Big(
{4\mu n^2 \over 3K }
+{n\over \xi^2}
\Big),
\label{}\end{align}
where the sum over discrete ${\mathbf{Q}}$'s has been replaced by integration ($\sum_{\mathbf{Q}} \approx n^{-1}\int d^2 {\mathbf{Q}}$). The term $\propto n$ represents a renormalization of the vortex self-energy, while the term $\propto n^2$ is the elasticity-driven interaction.

In the anisotropic case the sum over $\mathbf{Q}$'s in Eq. \eqref{sum_QQ} also yields a term $\propto n$, which renormalizes the vortex self-energy, and a term $\propto n^2$ which contributes to the elasticity-driven interaction between vortices. In this case, a dependence on the orientation of the VL with respect to the crystal axes is implicit in these two terms.

Taking into account that $|\Psi_s|^4b'\sim H_{c}^2(\Delta K /K)$ and $n H_c \xi^2\sim B/\kappa $, the elasticity-driven interaction can be estimated as
\begin{align}
F_{\rm int}\sim -{(1 + 4\ln\kappa)^2\over \kappa^2}{\Delta K\over K}B^2.
\label{}\end{align}
As we see, its order of magnitude coincides with that we obtained in Sec. \ref{Q_estim} from qualitative estimations, as well as with that of the exact results that we obtained in Sec. \ref{isotrooopic} for the isotropic case. Let us mention that omitting $\ln \kappa$ in these expressions, i.e. omitting the non-core contribution to the vortex-induced strain, they reproduce the previously reported results.\cite{Kogan95}

\subsection{$\boldsymbol{H \approx H_{c2}}$\label{Hc2}}

Let us now consider the VL's near $H_{c2}$. When doing so, it is convenient to use the conventional dimensionless units instead of those defined in Eqs. \eqref{dimless}. These conventional units can be obtained from Eqs. \eqref{dimless} by replacing $|\Psi_s|^2$ with $ -a/b$. 

Following Kogan \cite{Kogan75} one can easily obtain, now from the equations of equilibrium \eqref{EE},
the so-called Abrikosov identities (see also Ref. \cite{Abrikosov57,Kogan81,Abrikosov}). In presence of strain they read
\begin{subequations}\label{Abrikosov}\begin{gather}
H_z=H_0 - {\omega \over 2\kappa},\label{Abrikosov1} \\
{\kappa - H_0\over \kappa}\langle \omega \rangle +
{1-2\kappa^2 \over 2\kappa^2}\langle \omega^2\rangle -
\alpha_{ij}\langle u_{ij}\omega \rangle =0. 
\label{Abrikosov2}
\end{gather}\end{subequations}
Here $H_0$ is a constant and $\omega$ is the squared modulus of the function $\Psi$, which is solution of the linearized equations \eqref{GL1} and \eqref{GL2} ($\Psi = \sqrt{\omega} e^{i\chi}$).
Bearing in mind that the magnetic induction is $B = \langle H_z\rangle = H_0 -\langle\omega\rangle/(2\kappa)$, from Eqs. \eqref{Abrikosov} one can also obtain the following relationship:
\begin{align}
\langle \omega \rangle =
{2\kappa (\kappa - B)
\over
\widetilde \beta _A - \beta _e },
\label{omega}\end{align}
where $\beta_e = -2\kappa^2 \alpha_{ij}\langle u_{ij} \omega \rangle /\langle \omega \rangle^2$ and  
$\widetilde \beta _A = 1 + (2\kappa^2 - 1)\beta_A$, with $\beta_A=\langle\omega^2\rangle/\langle \omega \rangle^2$.

Bearing in mind that from the Abrikosov identity \eqref{Abrikosov1} it follows that $\langle H^2\rangle=B^2+(\langle \omega^2\rangle-\langle \omega \rangle^2)/(4\kappa^2)$, the free energy \eqref{master_F} can be rewritten as 
\begin{align}
F&=B^2 -
{\widetilde \beta_A \over 4\kappa^2}\langle \omega  \rangle^2 
+ {1\over 2}\lambda_{ijkl}\langle u_{ij}u_{kl}\rangle.
\label{Ff}\end{align}
When calculating the strain tensor $u_{ij}$ (now the total strain), we must take into account that $|\Psi|^2=\omega$. Thus, expressing $u_{ij}$ in the form \eqref{u_LP}, one finds that
\begin{subequations}\begin{align}
\epsilon_{ij}&=-\alpha_{kl}\lambda_{ijkl}^{-1}
\langle \omega\rangle,\nonumber \\
u_{i}(\mathbf q)&= iS_j(\mathbf q)G_{ji}(\mathbf q)\omega(\mathbf q),
\label{}\end{align}\end{subequations}
where $\omega(\mathbf q)$ represents the Fourier transform of the function $\omega$. In consequence, the last term of Eq. \eqref{Ff} is 
\begin{widetext}
\begin{align}
{1\over 2}\lambda_{ijkl}\langle u_{ij}u_{kl}\rangle&=
{1\over 2}\Big[
\lambda_{ijkl}\epsilon _{ij}\epsilon _{kl}
+\sum_{\mathbf q\not = 0}
G_{ij}^{-1}(\mathbf q)u_i(\mathbf q)u_j(-\mathbf q)\Big]\nonumber \\
&=-{1\over 2}\Big[
\lambda_{ijkl}\lambda_{mnkl}^{-1}\alpha_{mn}\epsilon _{ij}\langle\omega\rangle 
+i\sum_{\mathbf q\not = 0}
G_{ij}^{-1}(\mathbf q)G_{kj}(\mathbf q)S_k(\mathbf q)
u_i(\mathbf q)\omega(-\mathbf q)\Big]\nonumber \\
&=-{1\over 2 }\Big[
\alpha_{ij}\epsilon_{ij}\langle\omega\rangle
+i\sum_{\mathbf q\not = 0}
S_i(\mathbf q)u_i(\mathbf q)\omega(-\mathbf q)\Big]
=-{1\over 2}\alpha_{ij} \langle u_{ij}\omega \rangle={\beta_e \over 4\kappa^2}\langle \omega \rangle^2.
\end{align}
\end{widetext}
As a result, the free energy \eqref{Ff} is
\begin{align}
F=B^2 -
{\widetilde \beta_A -\beta_e \over 4\kappa^2}\langle \omega  \rangle^2 = 
B^2 -{(\kappa - B)^2 \over \widetilde \beta_A - \beta_e }.\label{F_H2_ani}\end{align}

Let us mention that the form of this expression for the free energy differs substantially from that reported by Miranovi\' c {\it et al.} in Ref. \cite{Miranovic95}. The elasticity-driven interaction term that one obtains from Eq. \eqref{F_H2_ani} is several times smaller than the corresponding one in Ref. \cite{Miranovic95}. The validity of Eq. \eqref{F_H2_ani} can be checked by noting that it reproduces the isotropic case [see expression (II) in Eq. \eqref{freeenattr}]. In contrast, the expression reported in Ref. \cite{Miranovic95} does not (see Sec. \ref{C_with_R}). The reason is that is it obtained from an expression analogous to Eq. \eqref{master_F}, but erroneous [Eq. (20) of Ref. \cite{Miranovic95}]. It reads
\begin{align}
F={1\over v} \int \left({H^2\over 8\pi} - {b\over 2}|\Psi|^4 
+\alpha_{ij}u_{ij}|\Psi|^2
+ {1\over 2}\lambda_{ijkl}u_{ij}u_{kl}\right) dv,
\end{align}
One can see here that the term $\alpha_{ij}u_{ij}|\Psi|^2$ is taken into account twice: one explicitly and another implicitly in the term $-{b}|\Psi|^2/2$ which arises as result of the integration by parts showed at the beginning of this Section.

\section{Concluding remarks}

We have revised the contribution to the VL energy which is due to the vortex-induced strains, showing that essential corrections in the previous calculations are needed. The most important one is connected with the fact that, in high-$\kappa$ superconductors, not only the vortex cores induce strains in a significant way. There also exists a significant contribution associated with the non-core regions which, in fact, the most important ones for the VL energies at low fields ($H\ll H_{c2}$). As a result of the proper inclusion of all strain sources, the strength of the elasticity-driven interaction between vortices increases by a factor up to $\sim \ln^2 \kappa$ compared with the previously reported ones. 

It is known since long ago that the observed correlations between VL's and crystal lattices in dirty superconductors cannot be explained without the elasticity-driven interaction between vortices.\cite{Ullmaier73} This interaction has been proved to be important in clean superconductors also. For example, the VL's observed in $\rm Nb Se_2$ do not correspond to the minimum of the London energy. In Ref.\cite{Kogan95} Kogan {\it et al.} showed that, however, the difference in the London energies of the two possible competing structures is smaller than the difference in the energies of the corresponding elasticity-driven interactions. As we have mentioned, Kogan {\it et al.} underestimated the elasticity-driven interaction between vortices because they assumed that only the vortex cores induce strain but, even doing so, they pointed out the importance of this interaction in $\rm NbSe_2$. This importance is increased as a result of the present work, what should be taken into account especially in those cases in which previous estimates of the above mentioned differences concluded that it was that of the London energy the most important one. 

$\rm V_3Si$ might provide an example in which the latter case takes place. In Ref.\cite{Kogan97} it was claimed that in $\rm V_3Si$ the contribution to the VL energy which is due to the (underestimated) elasticity-driven interactions between vortices can be neglected compared to that contribution due to the nonlocal corrections to the London energy. But bear in mind that (i) the order of magnitude of these two contributions is the same, as it was shown in Ref. \cite{Kogan95} considering the vortex cores as the only sources of strains, and (ii) the strength of the elasticity-driven interaction is considerably stronger than it was reported, as we have shown in this paper. So it is quite probable that in $\rm V_3Si$, as well as in other superconductors with large $\kappa$, this elasticity-driven interaction between vortices is not only comparable, but even more important than the nonlocal corrections to the London energy.

\vspace{-.625cm}

\begin{acknowledgments}

\vspace{-.25cm}

We acknowledge L.S. Froufe for useful discussions. A.P.L. was supported from the ESF programme Vortex Matter in Superconductors at Extreme Scales and Conditions (VORTEX) and by Comunidad de Madrid (07N/005/2002). S.A.M. was supported from the Russian Fund for Fundamental Research (Grant No. 00-02-17746). 

\end{acknowledgments}




\begin{thebibliography}{999}

\bibitem{Anderson62} P.W. Anderson, Phys. Rev. Lett. {\bf 9}, 309 (1962).

\bibitem{Kramer67} E.G. Kramer and C.L. Bauer, Philos. Mag. {\bf 15}, 1189 (1967).

\bibitem{Labush68} R. Labusch, Phys. Rev. {\bf 170}, 470 (1968). 

\bibitem{Galaiko68} V.P. Galaiko, Pris'ma Zh. Eksp. Teor. Fiz. {\bf 7}, 294 (1968) [JETP Lett. {\bf 7}, 230 (1968)].

\bibitem{Miyahara69} K. Miyahara {\it et. al.}, J. Phys. Soc. Jpn. {\bf 27}, 290 (1969).

\bibitem{Campbell72} A.M. Campbell and J.E. Evetts, Adv. Phys. {\bf 21}, 199 (1972), and the references therein.

\bibitem{Kronmuller73} H. Kronmüller and R. Schmucker, Phys. Status Solidi B {\bf 57}, 667 (1973); {\bf 74}, 261 (1976). 

\bibitem{Chudnovsky01} E.M. Chudnovsky, Phys. Rev. B {\bf 64}, 212503 (2001).

\bibitem{observed_correlation} B. Obst, Phys. Status Solidi B {\bf 45}, 467 (1971); J. Schelten, H. Ullmaier, and W. Schmatz, {\it ibid} {\bf 48}, 649 (1971); U. Essmann, Physica {\bf 55}, 83 (1971); H.F. Hess, C.A. Murray and J.V. Waszzak, Phys. Rev. Lett. {\bf 69}, 2138 (1992), Phys. Rev. B {\bf 50}, 16528 (1994); C.A. Bolle et al., Phys. Rev. Lett. {\bf 71}, 4039 (1993); P.L. Gammel et al., Phys. Rev. Lett. {\bf 72}, 278 (1994). 

\bibitem{Ullmaier73} H. Ullmaier, R. Zeller and P.H. Dederichs, Phys. Lett. {\bf 44A}, 331 (1973).

\bibitem{Kogan95} V.G. Kogan {\it et al}., Phys. Rev. B {\bf 51}, 15 344 (1995).

\bibitem{Miranovic95} P. Miranovi\'c, L. Dobrosavljevi\'c-Gruji\'c and V.G. Kogan, Phys. Rev. B {\bf 52}, 12 852 (1995).

\bibitem{Simanek91}E. $\rm \breve{S}$im\' anek, Phys. Lett. A {\bf 154}, 309 (1991); M. Coffey, Phys. Rev. B {\bf 49}, 9774 (1994); J. Low Temp. Phys. {\bf 96}, 81 (1994); {\bf 97}, 181 (1994). 

\bibitem{Abrikosov57} A.A. Abrikosov, Zh. Eksp. Teor. Fiz. {\bf 32}, 1442 (1957) [Sov. Phys.- JETP {\bf 5}, 1174 (1957)].

\bibitem{Abrikosov} A.A. Abrikosov, {\it Introduction to the Theory of Metals} (Noth-Holland, Amsterdam 1986).
 
\bibitem{de Gennes64} P.G. de Gennes and J. Matricon, Rev. Mod. Phys. {\bf 36}, 45 (1964).

\bibitem{Landau}L.D. Landau and E.M. Lifshitz, {\it Theory of Elasticity} (Pergamon, New York, 1986).

\bibitem{nota1} The corresponding parameter of smallness is $\Delta K /K \sim \alpha_{ij}\alpha_{kl}\lambda_{ijkl}^{-1}$, where $\Delta K$ is the change in the bulk modulus $K$ at the normal-superconducting transition.  

\bibitem{Larkin_Pikin} A.I. Larkin and S.A. Pikin, Zh. Eksp. Teor. Fiz. {\bf 56}, 1664 (1969) [Sov. Phys. JETP {\bf 29}, 891 (1969)].

\bibitem{Khachaturyan} A.G. Khachaturyan, {\it Theory of Structural Transformation in Solids} (Wiley, New York, 1983); C. Teodosiu, {\it Elastic Models of Crystal Defects} (Springer-Verlag, 1982). 

\bibitem{Friedel63}J. Friedel, P.G. de Gennes and J. Matricon, Appl. Phys. Letters {\bf 2}, 119 (1963); A.L. Fetter, Phys. Rev. {\bf 147}, 153 (1966).

\bibitem{Fetter69} A.L. Fetter and P. Hohenberg, in {\it Superconductivity}, edited by R.D. Parks (Marcel Dekker, Inc., New York, 1969), Vol. 2, pp. 817-923.

\bibitem{nota2} Note that, in principle,  $d^{-1}$, $\lambda_{L}^{-1}$, and $\xi^{-1}$ are characteristic $q$'s of the functions $h_{0,1}(\mathbf{q})$ ($d$ is the distance between vortices). Close to $H_{c1}$ they are such that $d^{-1} \ll \lambda_{L}^{-1} \ll \xi^{-1}$. For intermediate fields, however, $\lambda_{L}$ is not a characteristic length of the VL because of the overlap between vortices. So here we have $d^{-1} \ll \xi^{-1}$.

\bibitem{Kogan75} V.G. Kogan, J. Low Temp. Phys. {\bf 20}, 103 (1975).

\bibitem{Kogan81} V.G. Kogan and J.R. Clem, Phys. Rev. B {\bf 24}, 2497 (1981).

\bibitem{Kogan97} V.G. Kogan {\it et al}., Phys. Rev. Lett {\bf 79}, 741 (1997).

\end{thebibliography}
\end{document}